\documentclass[conference]{IEEEtran}
\usepackage{epsfig}
\usepackage[cmex10]{amsmath}
\usepackage{url}

\usepackage{multirow}
\usepackage{array}
\usepackage{footnote}

\begin{document}

\title{Addressing the P2P Bootstrap Problem for Small Overlay Networks}

\author{
\IEEEauthorblockN{David Isaac Wolinsky, Pierre St. Juste, P. Oscar Boykin,
and Renato Figueiredo}
\IEEEauthorblockA{Advanced Computing Information Systems Lab\\
University of Florida}
}

\maketitle

\begin{abstract}

Peer-to-Peer (P2P) overlays provide a framework for building distributed
applications consisting of few to many resources with features including
self-configuration, scalability, and resilience to node failures.  Such systems
have been successfully adopted in large-scale Internet services for content
delivery networks, file sharing, and data storage.  In small-scale systems,
they can be useful to address privacy concerns as well as support for network
applications that lack dedicated servers.  The bootstrap problem, finding an
existing peer in the overlay, remains a challenge to enabling these services
for small-scale P2P systems.  In large networks, the solution to the bootstrap
problem has been the use of dedicated services, though creating and maintaining
these systems requires expertise and resources, which constrain their
usefulness and make them unappealing for small-scale systems.

This paper surveys and summarizes requirements that allow peers potentially
constrained by network connectivity to bootstrap small-scale overlays through
the use of existing public overlays.  In order to support bootstrapping, a
public overlay must support the following requirements: a method for reflection
in order to obtain publicly reachable addresses, so peers behind network
address translators and firewalls can receive incoming connection requests;
communication relaying to share public addresses and communicate when direct
communication is not feasible; and rendezvous for discovering remote peers,
when the overlay lacks stable membership.  After presenting a survey of various
public overlays, we identify two overlays that match the requirements:  XMPP
overlays, such as Google Talk and Live Journal Talk, and Brunet, a structured
overlay based upon Symphony.  We present qualitative experiences with
prototypes that demonstrate the ability to bootstrap small-scale private
structured overlays from public Brunet or XMPP infrastructures.

\end{abstract}

\section{Introduction}

While P2P overlays provide a scalable, resilient, and self-configuring platform
for distributed applications, their adoption rate for use across the Internet
has been slow outside of large-scale systems, such as data distribution and
communication.  General use of decentralized, P2P applications targeting homes
and small/medium businesses (SMBs) has been limited in large part due to
difficulty in decentralized discovery of P2P systems --- the bootstrap problem
--- further inhibited by constrained network conditions due to firewalls and
NATs (network address translators).  While these environments could benefit
from P2P, many of these users lack the resources or expertise necessary to
bootstrap private\footnote{In the context of this paper, private implies that
the overlay's purpose is not for general use. Once established, such overlays
can support privacy in communication; however, overlay security is beyond the
scope of this paper.} P2P overlays particularly when the membership is unsteady
and across wide-area network environments where a significant amount of (or
all) peers may be unable to initiate direct communication with each other due
to firewalls and NATs.

Examples of large-scale P2P systems include Skype, BitTorrent, and Gnutella.
Skype is a voice over P2P system, whereas BitTorrent and Gnutella are used for
file sharing.  The bootstrapping in these systems typically relies on overlay
maintainers using high availability systems for bootstrapping, bundling their
connection information with the application for distribution.  When the
application is started, it uses these high availability servers to connect with
other peers in the system.  Alternatively, some services constantly crawl the
network and place peer lists on dedicated web sites. A new peer wishing to join
the network queries the web site and then attempts to connect to the peers on
that list.

In smaller-scale systems, P2P interests focus on decentralization.  For
example, users may desire to run an application at many distributed sites, but
the application lacks dedicated central servers to provide discovery or
rendezvous service for peers.  In contrast, dedicated, centralized P2P service
providers, such as LogMeIn's Hamachi, a P2P VPN, may collect usage data, which
the users may wish to remain private, or are not free for use.

Many applications make sense for small-scale overlay usage, including
multiplayer games, especially those that lack dedicated online services;
private data sharing; and distributed file systems.  Clearly, a small P2P
system could be bootstrapped by one or more users of the system running on
public addresses, distributing addresses out-of-band, instructing their peers
to add that address to their P2P application, and then initiate bootstrapping;
but these types of situations are an exception and not the norm.  Ultimately,
the users would be enhanced significantly through approaches that can make
decentralized bootstrapping transparent through minimal and intuitive
interaction with the P2P component.

\begin{figure*}[h!t!]
\centering
\epsfig{file=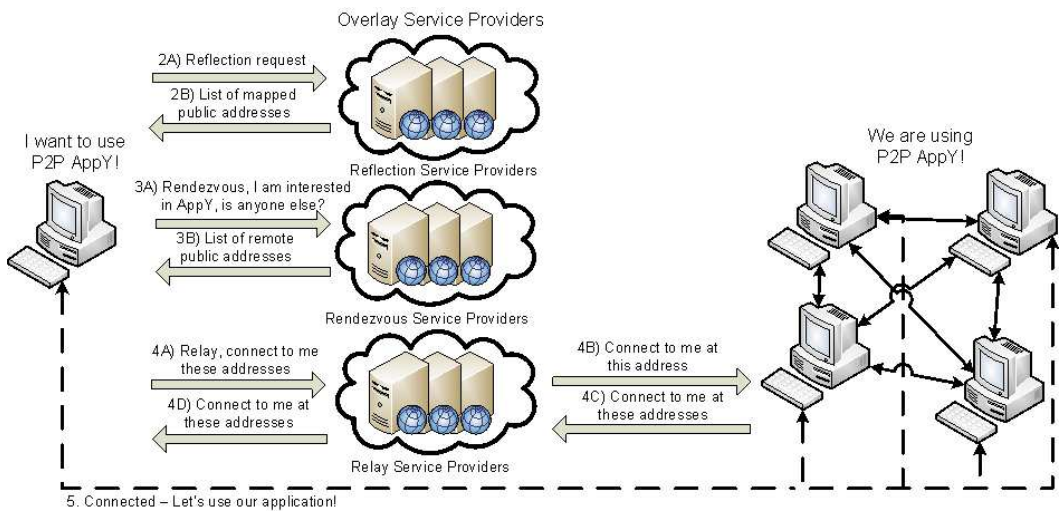, width=5.5in}
\caption{Bootstrapping a P2P system using an existing (generic) overlay.}
\label{fig:bootstrap}
\end{figure*}

The basic bootstrapping process can be broken down into two components: finding
a remote peer, and connecting to it and then successively more peers.  When a
node begins, it contacts various bootstrap servers, until it successfully
connects with one, upon which they exchange information.  The bootstrap server
may inquire into the overlay for the best set of peers for the new peer and
respond with that information or it may respond with its existing neighbor set.
At which point, the peer attempts to connect with those peers.  This process
continues aggressively until the peer arrives at a steady state, either
connecting with a specific set of or a number of peers.  Afterwards, the P2P
logic becomes passive, only reacting to churn from new incoming or outgoing
peers.

Overlay support for constrained peers, i.e., those behind NATs and restrictive
firewalls, requires additional features to support all-to-all connectivity for
peers in the overlay.  The instantiation of P2P systems for private use could
become overly burdensome, potentially relying on significant human interaction
to bootstrap them, for example, by relaying connection information through
phone calls and e-mail.  Even if this is feasible, this sort of interaction is
undesirable; P2P systems should be self-discovering so that users need to do
minimal amount of work to take advantage of them and ad-hoc systems stress this
point.  In addition, these may rely on centralized components; if they become
unavailable, which is a possibility since most users lack the expertise in
configuring highly available systems, the system will not be accessible.

To address this, we explore the use of existing public overlays as a means to
bootstrap private overlays.  There are many existing public overlays with high
availability, such as Skype, Gnutella, XMPP (Extensible Messaging and Presence
Protocol), and BitTorrent; by leveraging these systems, system integrators can
easily enable users to seamlessly bootstrap their own private P2P systems.  In
the preceding paragraphs, we identified the components necessary for
bootstrapping a homogeneous system; now we expand them for environments to
support the bootstrapping of a private overlay from a public overlay with
consideration for network constrained peers.  The public overlay must support
the following mechanisms as illustrated in Figure~\ref{fig:bootstrap}:
\begin{enumerate}
\item \textbf{Reflection} - A method for obtaining global application and
IP addresses or identifier for a peer that can be shared with others to enable
direct communication.
\item \textbf{Relaying} - A method for peers to exchange arbitrary data, when
a direct IP link is unavailable.
\item \textbf{Rendezvous} - A method for identifying peers interested in the
same P2P service.
\end{enumerate}
This work motivates from the belief that what prevents use of small-scale P2P
systems is due to lack the resources, technical knowledge, and lack of ability
and desire to create and manage high availability bootstrap services.  A public
overlay can be used to transparently bootstrap a private overlay with minimal
user interaction.

The requirements are presented and verified in the context of two prototype
implementations: a XMPP (Jabber)~\cite{xmpp} and Brunet~\cite{brunet}.
XMPP-based overlays are commonly used as chat portals, such as GoogleTalk and
Facebook Chat.  XMPP also supports an overlay amongst servers forming through
the XMPP Federation, which allows inter-domain communication amongst chat
peers, so that users from various XMPP servers can communicate with each other.
Brunet provides generic P2P abstractions as well as an implementation of the
Symphony structured overlay.  We present the architecture for these systems,
the lessons learned in constructing and evaluating them, and provide an
analysis of the latency to establish peer connectivity in a small-scale private
Brunet overlay with NAT-constrained nodes.

The organization of this paper follows.  Section~\ref{background} overviews
common P2P overlay technologies, motivating examples for this work, existing
solutions to the bootstrapping problem, and NAT challenges in P2P systems.  In
Section~\ref{overview}, we present a survey of overlays, applying the
requirements for private overlay bootstrapping to them, and then show in detail
how they can be applied to Brunet and XMPP.  Our implementation is described in
Section~\ref{implementation}.  In Section~\ref{evaluations}, we then perform a
timing evaluation of bootstrapping overlays using our prototype PlanetLab and
discuss experiences in deploying the system.  Finally, we conclude the paper
with Section~\ref{conclusions}.

\section{Motivation and Background}
\label{background}

The most well-known Internet P2P systems consist of a very large number of
nodes and users who benefit from the sheer scale of the system to accomplish
tasks such as sharing large files. There are also several applications that can
benefit from techniques developed for P2P systems to deliver features that are
desirable in small-scale systems, providing self-organizing frameworks upon
which applications can be built to support resource aggregation and
collaboration.  An example is found in virtual private networks (VPNs), which
provide end-to-end virtual network connectivity among trusted peers.

Private overlays enable truly decentralized, P2P VPNs.  A P2P VPN enables the
reuse of existing network applications on a P2P overlay without modifications.
The challenges of reflection, relaying, and rendezvous exist significantly in
these systems.  In centralized VPNs like OpenVPN, a dedicated server provides
for all three services.  P2P VPN solutions (such as Hamachi) provide similar
VPN functionality, with the improvement that peers can form direct connections
with each other which bypass the server.  A decentralized P2P VPN solution like
BrunetVPN~\cite{socialvpn} relies on a dedicated bootstrap overlay that runs
on PlanetLab.  Using the techniques described in this paper, BrunetVPN could
be extended so that it can be bootstrapped into private systems without
additional user configuration and relying on sustainable large-scale public
overlays.

Large-scale decentralized P2P architectures that can assist with the bootstrap
problem are unstructured and structured systems.  Unstructured
systems~\cite{gnutella, fasttrack} are generally constructed by peers
attempting to maintain a certain amount of connections to other peers in the
P2P system, whereas structured systems organize into well-defined topologies,
such as trees, 1-D rings, or hypercubes.  Though unstructured systems are
typically simpler to bootstrap and maintain, they rely on global knowledge,
flooding, or stochastic techniques to search for information in an overlay,
creating potential scalability constraints.  Alternatively, structured
systems~\cite{pastry, chord, symphony, kademlia, can} have guaranteed search
time typically with a lower bound of $O(\log N)$ and in some cases even
$O(1)$~\cite{beehive}.  The most common feature found in structured overlays is
the support for a decentralized storage / retrieval system called a distributed
hash table (DHT), that maps keys with associated data to specific node IDs in
an overlay.  

Another subset found of P2P systems are those that are not fully decentralized,
and deal with the bootstrap problem explicitly through centralized resources.
These include ``P2P VPNs'' like Hamachi, earlier file-sharing systems like the
original Napster, and tracker-based BitTorrent.  BitTorrent differentiates
itself by using the trackers as a gateway into the overlay; once inside, peers
exchange connection information with each other directly, relegating the
tracker as a fall back.  This approach has enabled BitTorrent to be modified to
support trackerless torrents through using a DHT.

\begin{figure*}[h!t!]
\centering
\epsfig{file=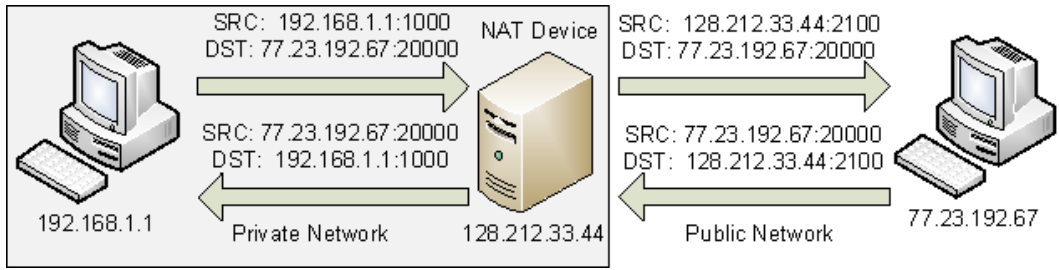, width=5.5in}
\caption{A typical NAT interaction. The peer behind a NAT has a private address.
When the packet is sent through the NAT, the NAT translates the source information
into a public mapping, keeping the original source information so that if a
packet from the remote peer comes back, it can be translated and delivered to
the original source.}
\label{fig:nat}
\end{figure*}

\subsection{Current Bootstrap Solutions}

As described in the introduction, the simple case of bootstrapping is limited
to one peer attempting to find an active peer in the overlay in order for
itself to become a member.  The large-scale providers have resources not
readily available to small-scale overlays.  This section reviews existing
techniques and those being developed and describes their application to
small-scale systems.

When using dedicated bootstrap overlays, a service provider hosts one or more
bootstrap resources.  Peers desiring to join the overlay query bootstrap nodes,
until a successful connection is made to one.  The bootstrap server will then
assist in connecting the peer to other nodes in the P2P system.  Bootstrap
nodes are either packaged with the application at distribution time or through
a meta data file, such as in BitTorrent.  Drawbacks to this approach for small,
ad-hoc pools include that the same server would have to be used every time to
bootstrap the system, or users would have to reconfigure their software to
connect to new bootstrap servers over time; at least one peer must have a
publicly accessible address; and a bootstrap server can become a single point
of failure.

Another commonly used approach for large-scale systems is the use of a host
cache~\cite{host_cache}.  Clients post current connection information to
dedicated web services, a host cache, that in turn communicate with other host
caches.  For small, ad-hoc networks, a host cache acts no differently than a
centralized rendezvous point, requiring that at least one peer has a publicly
accessible address.

``P2P VPN's''~\cite{p2pvpn} use of a BitTorrent tracker is similar to the host
cache concept.  The tracker hosts file meta data and peers involved in sharing.
For the VPN, the peer registers a virtual file used to organize the peers, a
form of rendezvous.  Each peer in the VPN queries the tracker regarding the
file, registers its IP address, and receives other active ``sharers'' IP
addresses.  Peers on public addresses or using UPNP are able to receive
incoming connections from all other peers.  The problem with this approach is
that it is heavily user-driven.  A user must register with each BitTorrent
tracker individually and maintain a connection with each of them, in order to
handle cases where BitTorrent trackers go offline.  In addition, this does not
use the BitTorrent trackers in a normal fashion, so it may be banned by tracker
hosts.

Research has shown that peers can use the locality properties of recent IP
addresses in a large-scale P2P system to make intelligent guesses about other
peers in the P2P system using an approach called random
probing~\cite{bootstrapping_p2p, locality_aware}.  The results show that, in a
network of tens to hundreds of thousands of peers, a bootstrapping peer can
find an active peer in 100 guesses to 2,000 guesses, depending on the overlay.
The approach does not really apply well to small-scale systems, especially when
peers are constrained by NATs and firewalls.

Rather than distribute an IP address, which points explicitly to some location
in the Internet, a small P2P network can apply a name abstraction around one
peer in the overlay using Dynamic DNS~\cite{bootstrapping_ddns}.  Peers share a
DNS entry, which points to a bootstrap server.  When the peers detect that the
bootstrap server is offline, at random time intervals they will update the DNS
entry with their own.  The application of this approach is well-suited to
small, ad-hoc groups, as the service could be distributed across multiple
Dynamic DNS registrations.  However, sharing a DNS entry requires trusting all
peers in the overlay, making it easy for malicious peers to inhibit system
bootstrapping.  Also the approach requires that at least one peer be publicly
addressable; if a non-publicly addressable peer updates the cache
inadvertently, it could delay or permanently prevent peers from creating a P2P
system.  The results reported in~\cite{bootstrapping_ddns} were
simulation-based and did not determine how well a dynamic DNS handles rapid
changing of name to IP mappings.

IP supports multicasting to groups interested in a common service.  In the case
of bootstrapping a P2P system~\cite{pastry, locality_aware}, all peers would be
members of a specific group.  When a new peer comes online, it queries the
group for connection information and connects to those that respond.  The
approach, by itself, requires that all peers are located in a multicast capable
network, restricting this approach typically to local area networks.

A large-scale structured overlay~\cite{one_ring, p2p_bootstrap} could enable
peers to publish their information into a dedicated location for their service
or application and then query that list to obtain a list of online peers.
Peers could search for other peers in their overlay and connect with them using
their connection information.  Since the service would be a large-scale system,
it could easily be bootstrapped by a dedicated bootstrap or host caches.  As it
stands, the described works were position papers and the systems have not been
fully fleshed out.  The primary challenge in relationship to small, ad-hoc
networks is that it lacks details bootstrapping of peers behind NATs into
overlays as it provides only a means for rendezvous and no reflection nor
relaying.

\subsection{NAT Hampering the Bootstrap Process}

As of 2010, the majority of the Internet is connected via Internet Protocol
(IP) version 4.  This protocol has a quickly approached limit of addresses
available,  only $2^{32}$ (approximately 4 billion).  With the Earth's
population at over 6.8 billion and each individual potentially having multiple
devices with Internet connectivity, the IPv4 limitation is becoming more and
more apparent.  Addressing this issue are two approaches:  1) the use of NATs
to enable many machines and devices to share a single IP address but preventing
bidirectional connection initiation, and 2) IPv6 which supports $2^{128}$
addresses.  The use of NATs, as shown in Figure~\ref{fig:nat}, complicates the
bootstrapping of P2P systems as it prevents peers from simply exchanging
addresses with each other to form connections, as the addresses may not be
public.  In addition, firewalls may prevent peers from receiving incoming
connections.  Thus, while the eventual widespread use IPv6 may eliminate the
need for address translation, it does not deal with the issue of firewalls
preventing P2P applications from communicating, and it is not clear that IPv6
users will not continue to rely on NAT/firewall devices to provide a
well-defined boundary of isolation for their local networks.

There are a handful of recognized NAT devices as presented in~\cite{stun,
p2p_nats_rfc}.  The following list focuses on the more prevalent types:
\begin{itemize}
\item \textbf{Full cone} - All requests from the same internal IP and port are
mapped to a static external IP and port, thus any external host can communicate
with the internal host once a mapping has been made.
\item \textbf{Restricted cone} - Like a full cone, but it requires that the
internal host has sent a message to the external host before the NAT will pass
the packets.
\item \textbf{Port restricted cone} - Like a restricted cone, but it requires
that the internal host has sent the packet to the external hosts specific port,
before the NAT will pass packets.
\item \textbf{Symmetric} - Each source and destination pair have no relation,
thus only a machine receiving a message from an internal host can send a
message back.
\end{itemize}

Because of the nature of NATs, two peers behind NATs will not be able to
communicate directly without each other without assistance.
Section~\ref{reflection} describes techniques that enable peers to acquire
routable addresses, Section~\ref{relay} describes relaying solutions, the
alternative when two peers are unable to communicate directly, and
Section~\ref{rendezvous} describes mechanisms for peers to identify each other
to exchange address and relaying information.

\section{Core Requirements}
\label{overview}

As presented in the preceding sections, a solution to bootstrapping small P2P
overlays must address several challenges, namely reflection, rendezvous, and
relaying.  In this section, we present a generic solution to this problem.  At
the basis of our solution is the use of a publicly available free-to-join
public overlay.  In order to support these features the public overlay must
have mechanisms for peers to obtain a public network identity (reflection);
search for other peers that are bootstrapping the same P2P service
(rendezvous); and send messages to peers through the overlay (relaying).  These
are the minimum requirements to bootstrap a decentralized, P2P system when all
peers are behind NATs.

\subsection{Reflection}
\label{reflection}

Reflection provides a peer with a globally-addressable identifier, which can be
shared with other peers so that it can receive incoming messages.  Without
reflection, peers on different networks with non-public addresses would not
have routable addresses with which to communicate with each other.  Reflection
does not need to be limited to IP.  For example, when a peer joins a service,
such as a chat application or a P2P system, the overlay provides a unique
identifier, which also serves as a form of reflection.

In IP communication, reflection can enable NAT traversal.  The simplest method
for NAT traversal is the multiplexing of a single UDP socket, IP address and
port combination.  This behavior can be supported through either local
configuration or remote assistance.  The local configuration approach relies on
the local router supporting either UPnP~\cite{upnp} or port forwarding /
tracking.  In many cases, UPnP is not enabled by default and in most commercial
venues it will rarely be enabled.  Port forwarding / tracking require a more
detailed configuration of a router, outside the comfort range of many
individuals and is not uniform across routers.  A peer using UPnP needs no
further services, as UPnP enables a peer to set and obtain both public IP
address and port mappings.  Port forwarding and tracking mechanisms still
require that the user obtains and inputs into the application their public IP
address or use in-band assistance described next.

In the remotely assisted scenario, a peer first sends a message to a reflection
provider, perhaps using STUN~\cite{stun_rfc}.  The response from the provider
tells the peer from which IP address and port the message was sent.  In the
case of all cone NATs, this will create a binding so that the peer can then
share that IP address and port with other peers behind NATs.  When the two
peers communicate simultaneously, all types of cone NATs can be traversed; the
timing of messages needs to be carefully considered, however, since NAT
mappings may change over time.  So long as one peer is behind a cone NAT, NAT
traversal using this mechanism is possible.  The situation becomes complicated
when both peers are behind symmetric NATs, or when either one of them have a
firewall prevent UDP communication.  

Peers behind symmetric NATs cannot easily communicate with each other, since
there is no relation between remote hosts and ports and local ports.  Further
complicating the matter is that there are various types of symmetric NATs,
having behaviors similar to the various cone NAT types. In~\cite{ice} the
authors describe methods to traverse these NATs so long as there is a
predictable pattern to port selection.  

Unlike UDP, TCP NAT traversal is complicated by the state associated with TCP.
In many systems, the socket API can be used to enable a peer to both listen for
incoming connections and form outgoing connections using the same local
addressing information.  According to~\cite{ice-tcp}, this method works for
various types of systems though the success rate on NATs is low, 40\%.  Other
mechanisms rely on out-of-band communication~\cite{pvc}, or use of complicated
predictive models~\cite{tcp-hole-punching}.

\subsection{Relaying}
\label{relay}

NAT traversal services only deal with one aspect of the bootstrap problem:
reflection.  That is, peers are able to obtain a public address for receiving
incoming connections.  They provide no means for users to exchange addresses
with each other or perform simultaneous open to traverse restrictive NATs.  To
address this issue, many systems incorporate these NAT traversal libraries and
use intermediaries to exchange addresses as a method of relaying.  Another form
of relaying exists when two peers are unable to form direct IP connections with
each other.

The most common method for relaying in IP is the use of TURN~\cite{turn}, or
Traversal Using Relay NAT.  A peer using TURN obtains a public IP address and
port that can be used as a forwarding address.  When a remote peer sends to
this address, the TURN server will forward the response to the peer who has
been allocated that mapping.  The lack of abstraction in TURN makes the system
heavily centralized, making its application in small-scale systems complicated.  

In overlays, peers typically have an abstracted identifier that does not
associate them with a single server enabling more decentralized approaches to
relaying.  When a remote peer sends a message to the identifier, the overlay
should translate the identifier into network level addresses and forward it to
the destination.

Relaying must, also, have reliability features like UDP enabling peers to
exchange various sized, arbitrary messages.  When a peer sends a message, it
should expect the remote peer should receive it in a reasonable amount of time
or not at all.  If the sending peer does not receive a response within a
reasonable amount of time, follow up requests can be sent until successful or
it is deemed the remote party is no longer online.  

Finally, the service should be asynchronous or event driven.  The previous
requirements would allow peers to relay through a message board or even by
posting messages to a DHT.  The problem with these two approaches is that peers
may very well communicate for long periods of time using these services.  That
means the potential for posting large amounts of data to a service that will
retain it and constantly querying the service to determine if an update is
available.  Both of these are highly undesirable and may be viewed as denial of
service or spam attacks.

\subsection{Rendezvous}
\label{rendezvous}

A rendezvous service should allow peers to discover peers interested in the
same service and provide a global identifier to contact that peer.  In the
simplest case, a peer could randomly probe other peers on the Internet until it
finds a matching peer.  This approach is unreasonable if overlay is small and
even more so if the peers are behind NATs, as the NATs may very well ignore the
requests even if a peer behind the NAT is actively looking for that overlay as
well.

Given an overlay, the most straightforward mechanism for rendezvousing is the
use of a broadcast query to determine if any other peers are using the same
service.  In small enough overlays, this is a perfectly reasonable approach,
though in large scale systems such as Gnutella, the approach is not scalable.

There is not one unified method to consolidate rendezvous as doing such would
severely limit its capabilities.  Programming rendezvous using the unique
features of an overlay can enable more efficient forms of rendezvous enabling
peers to increase the likelihood of finding a mutual peer and doing so more
quickly.  For example, in the case of a DHT, peers can use a single DHT key to
store multiple values, all of which would be addresses used to communicate with
peers in the overlay.  Alternatively, in a system like BitTorrent, peers could
use the same tracker and become ``seeds'' to the same virtual file.

\begin{table*}[h!t!]
\centering
\begin{tabular}[c]{|m{1.5cm}||m{5.5cm}|m{3cm}|m{3cm}|m{3cm}|}
\hline & Description & Reflection & Rendezvous & Relay \\ \hline \hline
BitTorrent &
Default BitTorrent implementations rely on a centralized tracker to provide the
initial bootstrapping.  Peers can establish new connections through information
obtained from established connections.  This relegates the tracker as a means
of monitoring the state of the file distribution.  BitTorrent specifies a
protocol, though each client may support additional features not covered by the
protocol.
&
The current specification does not support NAT traversal, though future
versions may potentially use UDP NAT traversal.  At which point, BitTorrent may
support a reflection service.
&
Peers can register as seeds to the same file hash, thus their IP address will
be stored with the tracker.
&
Peers receive each other's IP addresses from the tracker, there is no inherent
relaying.
\\ \hline
Gnutella &
Gnutella is a large-scale unstructured overlay with over a million peers;
primarily, it is used for file sharing.  Gnutella consists of a couple
hundred thousand ultra (super) peers to provide reliability to the overlay.
Gnutella is free-to-join and requires no registration to use.
&
Work in progress.  Peers attempt to connect to a sharer's resource, though a
"Push" notification reverses this behavior.  Thus a peer behind a NAT can
share with a peer on a public address.
&
Peers can perform broadcast searches with TTL up to 2; when networks consist of
millions of peers, small overlays will most likely not be able to discover each
other.
&
Not explicitly, could potentially utilize ping messages to exchange messages.
\\ \hline
Skype &
Skype is a large-scale unstructured overlay, consisting of over a million
active peers, and primarily used for voice over P2P communication.  Skype, like
Gnutella, also has super peers, though the owners of Skype provide
authentication and bootstrap servers.  Though Skype is free-to-join, it
requires registration to use.
&
Skype APIs provide no means for reflection.
&
Skype supports applications, or add-ons, which can used to transparently
broadcast queries to a users friend to determine if the peer has the
application installed.  Thus Skype does support rendezvous.
&
Skype applications are allowed to route messages via the Skype overlay, but
because Skype lacks reflection, all communication must traverse the Skype
overlay.
\\ \hline
XMPP &
XMPP consists of a federation of distributed servers.  Peers must register an
account with a server, though registration can be done through XMPP APIs without
user interaction.  XMPP is not a traditional P2P system, though it has some P2P
features.  XMPP servers on distinct servers are able to communicate with each
other.  Links between servers are created based upon client demand.  During link
creation, servers exchange XMPP Federation signed certificates.
&
While not provided by all XMPP servers, there exist extensions for NAT
traversal.  GoogleTalk, for example, provides both STUN and TURN servers.
&
Similar to Skype, XMPP friends can broadcast queries to each other to find
other peers using the same P2P service.  Thus XMPP supports rendezvous.
&
The XMPP specification allows peers to exchange arbitrary out-of-band
communication with each other.  Most servers support this behavior, even
when sent across the Federation.  Thus XMPP supports relaying.
\\ \hline
Kademlia~\cite{kademlia} &
There exists two popular Kademlia systems, one used by many BitTorrent systems,
Kad, and the other used by Gnutella, called Mojito.  Kademlia implements an
iterative structured overlays, where peers query each other directly when
searching the overlay.  Thus all resources of a Kademlia overlay must have
a publicly addressable network endpoint.
&
Existing implementations of Kademlia do not support mechanims for peers to
determine their network identity.
&
Peers can use the DHT as a rendezvous service, storing their connectivity
information in the DHT at key location:  $hash(SERVICE)$.
&
An iterative structured overlay has no support for relaying messages.
\\ \hline
OpenDHT~\cite{opendht} &
OpenDHT is a recently decommissioned DHT running on PlanetLab.  OpenDHT is
built using Bamboo, a Pastry-like protocol~\cite{pastry}.  Pastry implements
recursive routing, peers route messages through the overlay.
&
Existing implementations of Bamboo and Pastry do not support mechanims for
peers to determine their network identity.  Though this is ongoing work.
&
Peers can use the DHT as a rendezvous service, storing their connectivity
information in the DHT at key location:  $hash(SERVICE)$.
&
Because Pastry uses recursive routing, it can be used as a relay.  Furthermore,
extensions to Pastry have enabled explicit relays called virtual
connections~\cite{epost}.
\\ \hline
Brunet~\cite{brunet} &
Brunet like OpenDHT is a freely available DHT running on PlanetLab, though
still in active development.  Brunet creates a Symphony~\cite{symphony} overlay
using recursive routing.
&
Brunet supports inherent reflection services, when a peer forms a connection
with a remote peer, the peers exchange their view of each other.
&
Peers can use the DHT as a rendezvous service, storing their connectivity
information in the DHT at key location:  $hash(SERVICE)$.
&
Like Pastry, Brunet supports recursive routing and relays called
tunnels~\cite{hpdc08_0}.
\\ \hline
\end{tabular}
\caption{Public and Research Overlays}
\label{tab:overlays}
\end{table*}

\section{Implementations}
\label{implementation}

Table~\ref{tab:overlays} reviews various overlays, the majority of which are
high availability, public, free-to-join overlays, though some research only
overlays are included.  From this list, we chose to extend Brunet and XMPP
to support private overlay bootstrapping.  Brunet provides a structured P2P
infrastructure, though lacks an active, large-scale deployment outside of
academic institutions due to being rooted in an academic project.  XMPP, on the
other hand, enables connections between friends with routing occurring across a
distributed overlay.

Our implementation makes heavy use of the transports incorporated into
Brunet~\cite{brunet}.  The key distinguishing feature of this library is the
abstraction of sending over a communication link as it supports primitives
similar to ``send'' and ``receive'' that enables the ability to create P2P
communication channels over a variety of transports.  In the next sections, we
will describe how we extended Brunet to be self-bootstrapping as well as
extensions to enable bootstrapping from XMPP.

Our application of structured overlays as the basis private overlays focuses on
the autonomous, self-managing property of the overlay network rather than the
ability to scale to very large numbers.  This has also been the motivation of
related work which has employed structured overlays in systems in the order of
10s to 100s of nodes.  For example, Amazon's shopping cart runs on
Dynamo~\cite{dynamo} using a ``couple of hundred of nodes'' or less.  Facebook
provides an inbox search system using Cassandra~\cite{cassandra} running on
``600+ cores''.  Structured overlays simplify organization of an overlay and
provide each member a unique identifier abstracted from the underlying network.
As mentioned in the cited works, they provide high availability and autonomic
features that can handle churn well.  When used in small networks, most
structured overlays (including Brunet and Pastry) in effect act as $O(1)$
systems, self-organizing links that establish all-to-all connectivity among
peers.  Brunet explicitly supports all-to-all connectivity, though in some
cases may require constrained peers to route through relays.  This can further
be ensured by setting the amount of near connections for the infrastructures,
which in Brunet is configurable at run time.

The XMPP library we used is called Jabber-Net.  Each connection between peers
is uniquely identified by employing socket like concepts, i.e., a pair of
addresses and ports.  The basic representation for this constitutes a pair of
identifiers of the form ``brunet://P2P\_ID:PORT'', where each peer has a
unique ID and port associated for the local and remote entity.  The XMPP
implementation has a similar format: ``xmpp://USERNAME@DOMAIN:PORT/RESOURCE'',
again one identifier for the local peer and one for the remote.

\begin{figure*}[h!t!]
\centering
\epsfig{file=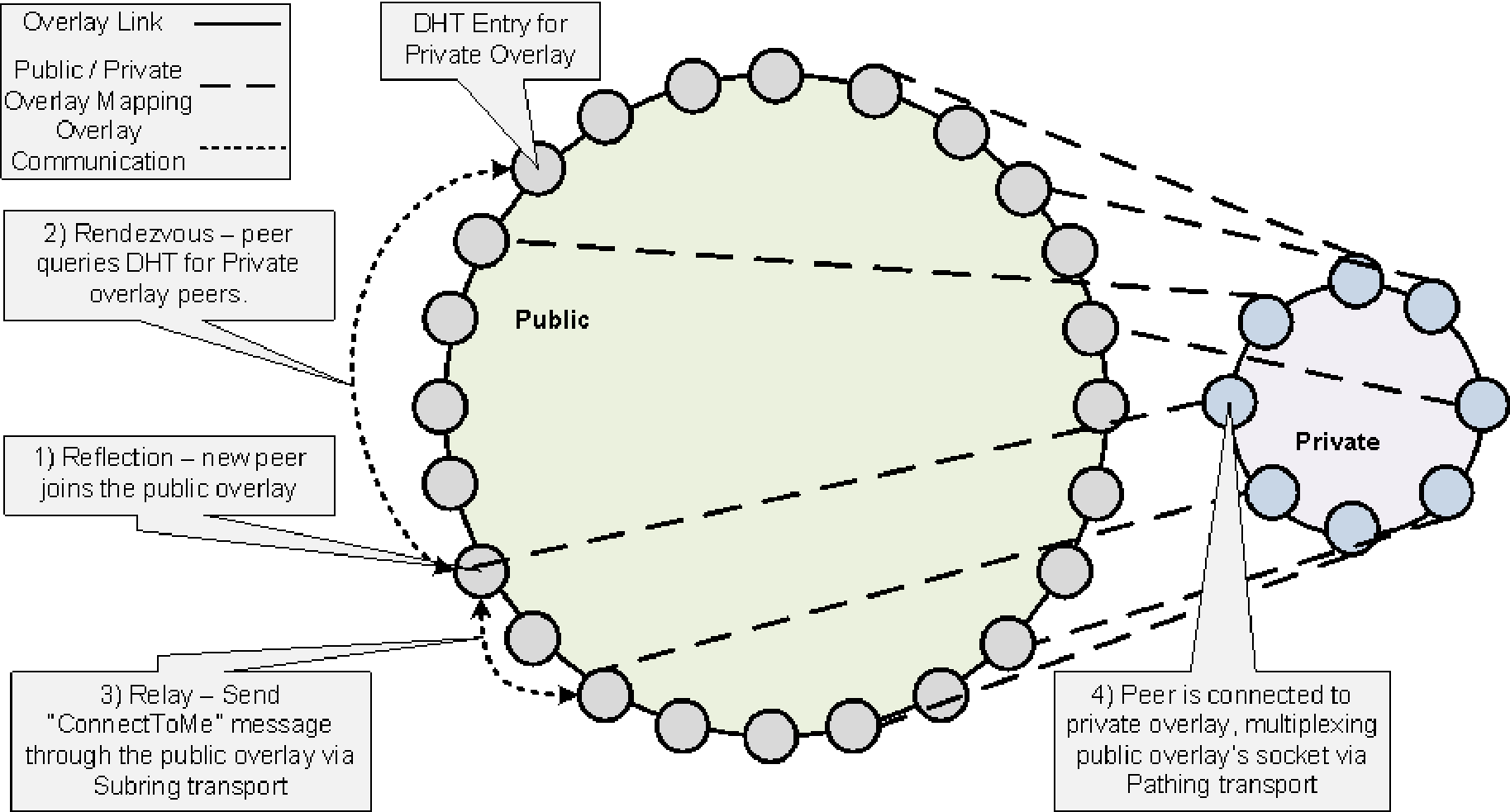, width=5.5in}
\caption{Bootstrapping a P2P system using Brunet.}
\label{fig:bootstrap_brunet}
\end{figure*}

\subsection{Bootstrapping Private Overlays Using Brunet}
\label{brunet_bootstrapping}

Prior to our work, Brunet bootstrapped using a recently online cache of peers
and IP multicast.  We have implemented Brunet to support STUN, such that, with
every connection Brunet makes, peers inform each other of their view of the
remote peers network state, a form of passive \textbf{reflection}.  Peers also
generate a unique 160-bit node identifier that can be used in the overlay as a
directly receive packets regardless of the underlay conditions.

In a single overlay, Brunet supports \textbf{relaying} either through the
overlay or pseudo direct connections called ``Tunnels''~\cite{hpdc08_0}, where
peers route to each other through common neighboring connections.  The relaying
in this context is used either to maintain a necessary overlay connection, or
to exchange intentions to connect with each other through ``ConnectToMe''
messages.  Thus when a peer desires a connection to another, both peers
simultaneously attempt to connect to each other after exchanging endpoints
discovered through reflection using the overlay relay mechanisms, dealing with
the issue of more restrictive cone NATs and the case when the peer is behind a
non-traversable NAT.  

To support \textbf{relaying} within the scope of a private overlay, we further
extended Brunet's transport library to support treating an existing overlay
as a medium for point-to-point communication. This is called a ``Subring''
transport, because it supports the abstraction of multiple private sub-rings
within a common large structured ring.  When the private overlay transmits data
across the public overlay, the private overlay packet is encapsulated (and
possibly encrypted) in a packet that ensures it will be delivered to the
correct private destination, and then encapsulated into a greedy routing
packet.  At which point, the packet is sent to the remote peer through the
public overlay.  In order to instruct peers to establish ``Subring'' links,
they exchange an identifier of the form ``brunet://P2P\_ID''.

Peers store their ``Subring'' identifiers into the DHT for \textbf{rendezvous}.
The DHT provides a scalable and self-maintaining mechanism for maintaining a
bootstrap, so long as the DHT supports multiple values at the same key, as
Brunet does.  The key used for the DHT rendezvous is a hash of the services
name and its version number, which we call a namespace.  Peers can then query
this entry in the DHT to obtain a list of peers in the private overlay.  Since
DHTs are soft-state, or lease systems, where data is released after a certain
period of time, an online peer must actively maintain its DHT entry.  In the
case that a peer goes offline, the DHT will automatically remove the value
after its lease has expired.

The final challenge faced was the application of Brunet's reflection service
for the private overlay.  There were two directions we could have gone.  The
first would have been to extend Brunet to support STUN in each of the remote
servers and then have a private node query them for their public information.
The problem with this approach is that it would require maintaining additional
state in order to discern which of the remote peers are on public addresses and
can provide STUN services.

Instead, we opted to multiplex the socket used for the public overlay as it
already had gone through the process of ``reflection''.  We call the
multiplexing of a single socket for multiple overlay ``Pathing''.  In this
context, the public and private overlays are given a virtual transport layer
that hooks into an underlying transport layer, thus not limited purely to
socket transport layers.  When peers exchange identifiers, instead of
transmitting a simple identifier like ``udp://192.168.1.1:15222'', the
``Pathing'' library extends it to ``udp://192.168.1.1:15222/path'', where each
path will signify a unique overlay.

Our completed approach is illustrated in Figure~\ref{fig:bootstrap_brunet}.
The approach of ``Subring'' and ``Pathing'' enabled the reuse of the core
components of Brunet.  Using ``Subring'' enables peers to form bootstrap
connections to then exchange ``ConnectToMe'' messages.  If the direct
connections failed, then the ``Subring'' connections could be used as permanent
connections.  The use of ``Pathing'' meant reuse of existing NAT traversal
techniques and limited the amount of system resources required to run multiple
overlays.  In terms of total lines of code, these abstractions enabled a
recursive overlay bootstrapping with a relatively small code footprint --- less
than 1000 lines of code.

\subsection{Bootstrapping Private Overlays Using XMPP}
\label{xmpp_bootstrapping}

In addition to supporting recursive bootstrapping of private overlays, the
techniques described above can be extended to use a different public overlay
--- an XMPP-based federation --- to support the bootstrapping of private
overlays.  The key features that make XMPP attractive are the distributed
nature of the federation and the openness of the protocol.  As of December
2009, there are over 70 active XMPP servers in the XMPP
Federation~\cite{xmpp_servers}.  These include GoogleTalk, Jabber.org, and Live
Journal Talk.

In XMPP, each user has a unique identifier of the form ``username@domain''.
Where the domain specifies the client XMPP server and the username explicitly
identifies a single individual.  XMPP supports concurrent instances for each
user by appending a resource identifier to the user ID:
``username@domain/resource''.  A resource identifier can either be provided by
the client or generated by the server.  For users in the same domain, the
server forwards the message from source to destination.  When two users are in
different domains, the sender's server forwards the message to the receiver's
server, who then relays it to the receiver.  Peers are able to send text
messages to each other as well as arbitrary binary messages called ``IQ''.

Peer relationships are maintained by the server.  Peer initiate them through an
in-band subscription mechanism based upon ``IQ'', allowing clients to handle
the process of adding and removing peers.  Once peers have established a
connection or subscription, they are informed through a ``Presence''
notification that the peer has come online, this include the full user
identifier.

The first form of \textbf{reflection} in XMPP is the unique client identifier.
Another is an IP reflection service available from some XMPP service providers
called ``Jingle''~\cite{jingle}.  ``Jingle'' uses ``IQ'' to determine available
STUN and TURN servers.  Fortunately, these services are provided free of charge
through GoogleTalk.  In Brunet, we extended the UDP transport to support
querying STUN servers so that it can obtain an address mapping and keep it
open.  STUNs protocol sets the first two bits to 0 in all messages, thus we
used that and the STUN cookie to distinguish it from other messages.

In order to support the situation where two peers are unable to communicate
through the exchanged addresses, we have extended XMPP ``IQ'' as a transport to
support \textbf{relaying}.  Once a peer has formed a connection through XMPP,
they are then able to attempt simultaneous connection attempts, in the same
fashion as the ``Subring'' transport, further increasing the likelihood of a
connection.  If that does not succeed, the peers can still relay through XMPP.
This approach also has the benefit that, if a XMPP server does not support
``Jingle'', the two peers can still form links with each other.  Since Brunet
internally supports IP reflection, eventually, if one of the peers in the system
has a public address, it will automatically assist the other peers into forming
direct links with each other.

\textbf{Rendezvous} uses a two step approach.  First peers advertise their use
of private overlay in the resource identifier.  The name is hashed to ensure
that the users complete identifier does not extend past 1,023 bytes, the
maximum length for these identifiers.  In addition, a cryptographyically
generated random number is appended to the resource identifier to distinguish
between multiple instances of the users application in the same private
overlay.  Once a peer receives a presence notification from a remote peer and
the base components match, that is the hash of the service, the peer adds it to
a list of known online peers.  If the peer lacks connections, the system
broadcasts to that list a request for addresses.  The peers respond with a list
of addresses including UDP, TCP, and XMPP addresses, concluding rendezvous.

Ideally, peers would not need to create XMPP connections with each other; if
they are on a public address, the rendezvous phase alone will suffice.  In the
case that they are not on public addresses, peers can first obtain their public
address through STUN, then form an XMPP connection with each other, and finally
perform simultaneous connection attempts.  If NAT traversal fails, the peers
can continue routing through the XMPP connection.  Due to the abstractions
employed by the transport library, the additional support for XMPP-based
bootstrapping required only an additional 700 lines of code to Brunet and no
modification to the core system.

\section{Evaluating Overlay Bootstrapping}
\label{evaluations}
In this section, we present a qualitative evaluation of our prototype
bootstrapping a small-scale network and also share some experiences with
deploying overlays.

\subsection{Deployment Experiments}

The purpose of our experiment is to verify that our techniques work and what
overheads should be expected in using Brunet and XMPP to bootstrap an
overlay.  Rather than an extensive experiment overly focused on overheads of
Brunet and XMPP, this experiment is primarily focused on the feasibility of
forming small-scale overlays among network-constrained peers.  The experiment
represents 5 peers desiring all-to-all direct connectivity, a feature
transparently available to them if they bootstrap into a private Brunet
overlay. The experiments were run on peers deployed on 5 distinct virtual
machines --- each virtual machine had its own separate NAT, and thus peers were
unable to communicate directly without assistance.

The public Brunet overlay used in this experiment consisted of over 600 nodes
and ran on PlanetLab.  PlanetLab~\cite{planetlab} is a consortium of research
institutes sharing hundreds of globally distributed network and computing
resources.  GoogleTalk provided the XMPP overlay used in this experiment.
Though this experiment does not take into advantage the features of the XMPP
Federation, this aspect is presented in more detail in the next section
reviewing experiences deploying overlays using XMPP.

In the experiment, 5 P2P nodes were started simultaneously, while measuring the
time spent for reflection, rendezvous, reflection, and connection.  The results
are presented in Table~\ref{tab:results}.  For XMPP, these are translated as
follows:  reflection measures the time to obtain IP addresses from the STUN
server, rendezvous is the time to receive a presence notification, relaying is
the time to receive a message across XMPP, and connected is once all nodes in
the private overlay has all-to-all connectivity.  For Brunet, these are
translated as follows:  reflection measures the time to connect to the public
overlay, rendezvous is the time to query the DHT, relaying is the average time
to send a message across the overlay, and connected is the time until the
private overlay has all-to-all connectivity.    The results are highly
correlated to timeouts in Brunet, which employs a mixture of events and
polling to stabilize the overlay, as well as the latency between the client and
GoogleTalk.  As this was more of a qualitative experiment, the results are
clear: private overlays providing all-to-all connectivity among NATed nodes can
bootstrap within a very reasonable amount of time.

\begin{table}[ht]
\centering
\begin{tabular}{|c||c|c|c|c|}
\hline & Reflection & Rendezvous & Relaying & Connected \\ \hline \hline
XMPP & .035 & .110 & .243 & 20.3 \\ \hline
Brunet & 3.05 & .330 & .533 & 23.22 \\ \hline
\end{tabular}
\caption{Time in seconds for various private overlay operations}
\label{tab:results}
\end{table}

\subsection{Deployment Experiences}

Recently, Facebook announced that they would be supporting XMPP as a means to
connect into Facebook chat.  This was rather exciting and further motivated
this work, as Facebook has over 400 million active users, which would have made
their XMPP overlay, potentially, the largest free-to-join overlay.
Unfortunately, Facebook does not employ a traditional XMPP setup, instead it
provides a proxy into their chat network, preventing features like arbitrary
IQs and other forms of out-of-band messages to be exchanged between peers.
User identifiers are also translated, so a peer cannot obtain a remote peers
real identifier.  Thus there exists no out-of-band mechanism for rendezvous.
Peers could potentially send rendezvous messages through the in-band XMPP
messaging, but this may be viewed by most recipients as spam as it would arrive
as normal chat messages.  The lesson learned was that XMPP servers not
associated with the Federation will not necessarily support features necessary
to bootstrap.

During initial tests in verifying the workings of the XMPP code base, we
bootstrapped a  private Brunet overlay on PlanetLab through various XMPP
service providers.  We discovered that some servers were ignoring clients on
PlanetLab.  Another server crashed after 257 concurrent instances of the same
account logged in.  Unfortunately, the provider had no contact information
available, so we were unable to determine if our test caused the crash.  Though
there did exist some servers that had no trouble hosting over 600 concurrent
instances running on PlanetLab.

Once the system was running on PlanetLab, more tests were performed to
determine the ability to bootstrap across the XMPP Federation.  We formed
friendships, or subscriptions, between users across a few different XMPP
service providers.  In the most evaluated case, a single peer on GoogleTalk
along with 600 peers on PlanetLab system using \textit{jabber.rootbash.com},
the GoogleTalk peer would not always receive presence notifications for all
peers online, though always would receive some.  When a peer began the relaying
mechanism, it would broadcast to every peer from whom it received a presence
notification.  When performing this between GoogleTalk and \textit{rootbash},
the GoogleTalk peer would not receive a response.  Though in reducing the
broadcast to a random selection of 10 peers, every 10 seconds until the
GoogleTalk peer was connected, the peer received responses.  The behavior
indicates that the XMPP servers may have been filtering to prevent denial of
service attacks.

Peers on the same XMPP server seem to be connected very quickly, though peers
on different services can take significantly longer.  For example, when
bootstrapping a single peer from GoogleTalk into the \textit{rootbash} system,
it always took 1 minute for the node to become fully connected to the private
overlay.  When the peer used \textit{rootbash}, the peer always connected
within 30 seconds.  It seems as if the communication between XMPP servers was
being delayed for some reason.  The same behavior was not experienced, when
chatting between the two peers.

\section{Conclusion}
\label{conclusions}

In this paper, we have established the requirements for bootstrapping
small-scale P2P overlays: reflection, relaying, and rendezvous.  Reflection is
required so that peers behind NATs and firewalls can obtain public addresses to
share with remote peers.  Relaying provides a means for peers to coordinate
simultaneous connection attempts and a fall-back in the case that NAT traversal
attempts fail.  Rendezvous is a common problem, even for large-scale systems;
peers must have a mechanism to find other peers connecting to are are in the
same overlay.  

We extended Brunet to support bootstrapping small-scale overlays from
another Brunet overlay as well as XMPP.  XMPP was chosen as readily
supported the key features required for bootstrapping small-scale P2P overlays.
The Brunet solution offers interesting insights. In particular, we have
found that the ability to multiplex a single socket or transport for multiple
overlays is important in practice for overlays that use UDP as a transport and
support NAT traversal, because a peer does not need to actively maintain
multiple NAT mappings.  XMPP provides a reliable, production ready, mechanism
of bootstrapping overlays as the service is hosted by a distributed set of
providers, each offering interoperability.  Our prototype clearly identifies
methods for reflection, relaying, and rendezvous and can assist in discovering
methods for doing so in similar systems.

For future work, we plan on investigating how peers can leverage existing DHT
deployments, such as Kad or Mojito, for rendezvous, form friendships
automatically in XMPP, and continue the bootstrap process using XMPP framework
discussed in this paper.  Using this pproach would enable the XMPP solution to
be extended to bootstrap overlays that are not based on social connections
alone.


\bibliographystyle{IEEEtran}
\bibliography{VirtualPrivateOverlays}

\begin{thebibliography}{10}
\providecommand{\url}[1]{#1}
\csname url@samestyle\endcsname
\providecommand{\newblock}{\relax}
\providecommand{\bibinfo}[2]{#2}
\providecommand{\BIBentrySTDinterwordspacing}{\spaceskip=0pt\relax}
\providecommand{\BIBentryALTinterwordstretchfactor}{4}
\providecommand{\BIBentryALTinterwordspacing}{\spaceskip=\fontdimen2\font plus
\BIBentryALTinterwordstretchfactor\fontdimen3\font minus
  \fontdimen4\font\relax}
\providecommand{\BIBforeignlanguage}[2]{{%
\expandafter\ifx\csname l@#1\endcsname\relax
\typeout{** WARNING: IEEEtran.bst: No hyphenation pattern has been}%
\typeout{** loaded for the language `#1'. Using the pattern for}%
\typeout{** the default language instead.}%
\else
\language=\csname l@#1\endcsname
\fi
#2}}
\providecommand{\BIBdecl}{\relax}
\BIBdecl

\bibitem{xmpp}
P.~Saint-Andre, ``{RFC 3920} extensible messaging and presence protocol
  ({XMPP}): Core,'' October 2004.

\bibitem{brunet}
P.~O. Boykin and et~al., ``A symphony conducted by brunet,''
  \url{http://arxiv.org/abs/0709.4048}, 2007.

\bibitem{socialvpn}
P.~S. Juste, D.~Wolinsky, P.~O. Boykin, M.~J. Covington, and R.~J. Figueiredo,
  ``Socialvpn: Enabling wide-area collaboration with integrated social and
  overlay networks,'' 2010.

\bibitem{gnutella}
T.~Klingberg and R.~Manfredi, ``Gnutella 0.6,''
  \url{http://rfc-gnutella.sourceforge.net/src/rfc-0_6-draft.html}, June 2002.

\bibitem{fasttrack}
hex, ``The fasttrack protocol,''
  \url{http://cvs.berlios.de/cgi-bin/viewcvs.cgi/gift-fasttrack/giFT-FastTrack%
/PROTOCOL}, September 2004.

\bibitem{pastry}
A.~Rowstron and P.~Druschel, ``Pastry: {Scalable,} decentralized object
  location and routing for large-scale peer-to-peer systems,'' in
  \emph{International Conference on Distributed Systems Platforms
  (Middleware)}, November 2001.

\bibitem{chord}
I.~Stoica and et~al., ``Chord: {A} scalable {Peer-To-Peer} lookup service for
  internet applications,'' in \emph{SIGCOMM}, 2001.

\bibitem{symphony}
G.~S. Manku, M.~Bawa, and P.~Raghavan, ``Symphony: distributed hashing in a
  small world,'' in \emph{USITS}, 2003.

\bibitem{kademlia}
P.~Maymounkov and D.~Mazi\`{e}res, ``Kademlia: A peer-to-peer information
  system based on the {XOR} metric,'' in \emph{IPTPS '02}, 2002.

\bibitem{can}
S.~Ratnasamy, P.~Francis, S.~Shenker, and M.~Handley, ``A scalable
  content-addressable network,'' in \emph{SIGCOMM}, 2001.

\bibitem{beehive}
V.~Ramasubramanian and E.~G. Sirer, ``Beehive: O(1)lookup performance for
  power-law query distributions in peer-to-peer overlays,'' in \emph{Symposium
  on Networked Systems Design and Implementation}, 2004.

\bibitem{host_cache}
H.~Damfpling. (2003) Gnutella web caching system.
  \url{http://www.gnucleus.com/gwebcache/specs.html}.

\bibitem{p2pvpn}
W.~Ginolas. (2009) {P2PVPN}. \url{http://p2pvpn.org}.

\bibitem{bootstrapping_p2p}
C.~GauthierDickey and C.~Grothoff, ``Bootstrapping of peer-to-peer networks,''
  2008.

\bibitem{locality_aware}
C.~Cramer, K.~Kutzner, and T.~Fuhrmann, ``Bootstrapping locality-aware p2p
  networks,'' in \emph{in: The IEEE International Conference on Networks
  (ICON)}, 2004.

\bibitem{bootstrapping_ddns}
M.~Knoll, A.~Wacker, G.~Schiele, and T.~Weis, ``Bootstrapping in peer-to-peer
  systems,'' 2008.

\bibitem{one_ring}
M.~Castro, P.~Druschel, A.-M. Kermarrec, and A.~Rowstron, ``One ring to rule
  them all: {Service} discover and binding in structured peer-to-peer overlay
  networks,'' in \emph{SIGOPS European Workshop}, Sep. 2002.

\bibitem{p2p_bootstrap}
M.~Conrad and H.-J. Hof, ``A generic, self-organizing, and distributed
  bootstrap service for peer-to-peer networks,'' in \emph{International
  Workshop on Self-Organizing Systems (IWSOS)}, 2007.

\bibitem{stun}
J.~Rosenberg, J.~Weinberger, C.~Huitema, and R.~Mahy. (2003) Stun - simple
  traversal of user datagram protocol (udp) through network address translators
  (nats).

\bibitem{p2p_nats_rfc}
P.~Srisuresh, B.~Ford, and D.~Kegel, \emph{{RFC} 5128 State of Peer-to-Peer
  ({P2P}) Communication across Network Address Translators ({NATs})}, March
  2008.

\bibitem{upnp}
``{UPnP} device architecture 1.1,''
  \url{http://www.upnp.org/specs/arch/UPnP-arch-DeviceArchitecture-v1.1.pdf},
  October 2008.

\bibitem{stun_rfc}
J.~Rosenberg, R.~Mahy, P.~Matthews, and D.~Wing, ``{RFC 3489} session traversal
  utilities for nat ({STUN}),'' October 2008.

\bibitem{ice}
J.~Rosenberg, ``Interactive connectivity establishment ({ICE}): A protocol for
  network address translator ({NAT}) traversal for offer/answer protocols,''
  \url{http://tools.ietf.org/html/draft-ietf-mmusic-ice-19}, October 2008.

\bibitem{ice-tcp}
S.~Perreault and J.~Rosenberg, ``{TCP} candidates with interactive connectivity
  establishment ({ICE}),''
  \url{http://tools.ietf.org/html/draft-ietf-mmusic-ice-tcp-08}, October 2009.

\bibitem{pvc}
A.~Rezmerita, T.~Morlier, V.~Neri, and F.~Cappello, ``Private virtual cluster:
  Infrastructure and protocol for instant grids,'' in \emph{Euro-Par}, November
  2006.

\bibitem{tcp-hole-punching}
A.~Biggadike, D.~Ferullo, G.~Wilson, and A.~Perrig, ``{NATBLASTER}:
  Establishing {TCP} connections between hosts behind {NAT}s,'' in \emph{ACM
  SIGCOMM Asia Workshop}, April 2005.

\bibitem{turn}
J.~Rosenberg, R.~Mahy, and P.~Matthews. (2009) "traversal using relays around
  nat (turn)". \url{http://tools.ietf.org/html/draft-ietf-behave-turn-16}.

\bibitem{opendht}
S.~Rhea and et~al., ``Opendht: a public dht service and its uses,'' in
  \emph{SIGCOMM}, 2005.

\bibitem{epost}
A.~Mislove, A.~Post, A.~Haeberlen, and P.~Druschel, ``Experiences in building
  and operating epost, a reliable peer-to-peer application,'' in \emph{EuroSys
  '06: Proceedings of the 1st ACM SIGOPS/EuroSys European Conference on
  Computer Systems 2006}, 2006.

\bibitem{hpdc08_0}
A.~Ganguly and et~al., ``Improving peer connectivity in wide-area overlays of
  virtual workstations,'' in \emph{HPDC}, 2008.

\bibitem{dynamo}
G.~DeCandia, D.~Hastorun, M.~Jampani, G.~Kakulapati, A.~Lakshman, A.~Pilchin,
  S.~Sivasubramanian, P.~Vosshall, and W.~Vogels, ``Dynamo: amazon's highly
  available key-value store,'' in \emph{SOSP '07: Proceedings of twenty-first
  ACM SIGOPS symposium on Operating systems principles}.\hskip 1em plus 0.5em
  minus 0.4em\relax New York, NY, USA: ACM, 2007, pp. 205--220.

\bibitem{cassandra}
A.~Lakshman, ``Cassandra – a structured storage system on a {P2P} network,''
  \url{http://www.facebook.com/note.php?note_id=24413138919}, August 2008.

\bibitem{xmpp_servers}
``Public {XMPP} services,'' \url{http://xmpp.org/services/}, December 2009.

\bibitem{jingle}
S.~Ludwig and et~al., ``{XEP}-0166: Jingle,'' December 2009.

\bibitem{planetlab}
B.~Chun, D.~Culler, T.~Roscoe, A.~Bavier, L.~Peterson, M.~Wawrzoniak, and
  M.~Bowman, ``Planetlab: an overlay testbed for broad-coverage services,''
  \emph{SIGCOMM}, 2003.

\end{thebibliography}
\suppressfloats

\end{document}